\documentclass[%
prb,%
sd,%
amsmath,amssymb,
reprint,%
superscriptaddress,
]{revtex4-1}
\usepackage{graphicx}% Include figure files
\usepackage{dcolumn}% Align table columns on decimal point
\usepackage{bm}% bold mat
\usepackage{textgreek}
\usepackage{color,soul}% mark text by wrapping it with \hl{}
\usepackage{comment}
\usepackage{multirow}%use for multirows in tables

\begin{document}
	
	\preprint{AIP/123-QED}
	
	\title[Magnetic properties and domain structure of ultrathin YIG/Pt bilayers]{Magnetic properties and domain structure \\ of ultrathin yttrium iron garnet/Pt bilayers}% Force line breaks with \\
	
	\author{J.~Mendil}
	\affiliation{Department of Materials, ETH Zurich, CH-8093 Zurich, Switzerland}
	
	\author{M.~Trassin}
	\affiliation{Department of Materials, ETH Zurich, CH-8093 Zurich, Switzerland}
	
	\author{Q.~Bu}
	\affiliation{Department of Materials, ETH Zurich, CH-8093 Zurich, Switzerland}
	
	\author{J.~Schaab}
	\affiliation{Department of Materials, ETH Zurich, CH-8093 Zurich, Switzerland}
	
	\author{M.~Baumgartner}
	\affiliation{Department of Materials, ETH Zurich, CH-8093 Zurich, Switzerland}
	
	\author{C.~Murer}
	\affiliation{Department of Materials, ETH Zurich, CH-8093 Zurich, Switzerland}
	
	\author{P. T.~Dao}
	\affiliation{Department of Materials, ETH Zurich, CH-8093 Zurich, Switzerland}
	
	\author{J. Vijayakumar}
	\affiliation{Swiss Light Source, Paul Scherrer Institut, CH-5232 Villigen PSI, Switzerland}
	
	\author{D. Bracher}
	\affiliation{Swiss Light Source, Paul Scherrer Institut, CH-5232 Villigen PSI, Switzerland}
	
	\author{C.~Bouillet}
	\affiliation{ Institut de Physique et Chimie des Mat\'{e}riaux de Strasbourg (IPCMS), UMR 7504 CNRS, Universit\'{e} de Strasbourg, 67034 Strasbourg, France}
	
	\author{C.~A.~F.~Vaz}
	\affiliation{Swiss Light Source, Paul Scherrer Institut, CH-5232 Villigen PSI, Switzerland}
	
	\author{M.~Fiebig}
	\affiliation{Department of Materials, ETH Zurich, CH-8093 Zurich, Switzerland}
	
	\author{P.~Gambardella}
	\affiliation{Department of Materials, ETH Zurich, CH-8093 Zurich, Switzerland}
	
	\date{\today}
	
\begin{abstract}
We report on the structure, magnetization, magnetic anisotropy, and domain morphology of ultrathin yttrium iron garnet (YIG)/Pt films with thickness ranging from 3 to 90~nm. We find that the saturation magnetization is close to the bulk value in the thickest films and decreases towards low thickness with a strong reduction below 10~nm. We characterize the magnetic anisotropy by measuring the transverse spin Hall magnetoresistance as a function of applied field. Our results reveal strong easy plane anisotropy fields of the order of 50-100~mT, which add to the demagnetizing field, as well as weaker in-plane uniaxial anisotropy ranging from 10 to 100 \textmugreek T. The in-plane easy axis direction changes with thickness, but presents also significant fluctuations among samples with the same thickness grown on the same substrate. X-ray photoelectron emission microscopy reveals the formation of zigzag magnetic domains in YIG films thicker than 10~nm, which have dimensions larger than several 100~\textmugreek m and are separated by achiral N\'{e}el-type domain walls. Smaller domains characterized by interspersed elongated features are found in YIG films thinner than 10~nm.
		
\end{abstract}
\maketitle
	
	\section{Introduction}
	Yttrium iron garnet (YIG) thin films have attracted considerable interest in the field of spintronics due to the possibility of converting magnon excitations into spin and charge currents flowing in an adjacent nonmagnetic metal (NM) layer. Spin currents in YIG/NM bilayers have been excited thermally (spin Seebeck effect) \cite{uchida2010observation,schreier2013current,wang2014joule,vlietstra2014simultaneous}, dynamically (spin-pumping) \cite{heinrich2011spin,castel2012frequency,wang2013large,castel2014yttrium} or by means of the spin Hall effect\cite{cornelissen2015long,cornelissen2016magnetic,cornelissen2016magnon,emori2018spin,PhysRevB.92.174406}. In the latter case, a charge current in the NM generates a transverse spin current that is either absorbed or reflected at the interface with YIG. This leads to a variety of interesting effects such as the spin Hall magnetoresistance\cite{nakayama2013spin,chen2013theory,hahn2013comparative,althammer2013quantitative} (SMR) and current-induced spin-orbit torques\cite{vlietstra2013exchange,schreier2015current,fang2017thickness}, which can be used to sense and manipulate the magnetization. For the latter purpose, it is desirable to work with thin magnetic films in order to achieve the largest effect from the interfacial torques.

For a long time, the growth of YIG has been accomplished by liquid phase epitaxy, which offers excellent epitaxial quality and dynamic properties such as low damping and a rich spin-wave spectrum. The magnetic properties of these bulk-like samples, including the magnetocrystalline anisotropy\cite{pearson1962magnetocrystalline,hansen1978rare} and magnetic domain structure\cite{heinz1971mobile,thiaville1988direct,Lisovskii2013}, have been extensively characterized in the past. However, with rare exceptions \cite{pirro2014spin}, samples grown by liquid phase epitaxy usually have thicknesses in the \textmugreek m to mm range. Recent developments in oxide thin film growth give access to the sub-\textmugreek m range by employing techniques such as laser molecular beam epitaxy \cite{krichevtsov2017magnetization}, sputtering \cite{Wang2014,Cooper2017}, and pulsed laser deposition (PLD) \cite{dorsey1993epitaxial,manuilov2009pulsed,manuilov2010pulsed,sun2012growth,d2013inverse,onbasli2014pulsed,sokolov2016thin,hauser2016yttrium}, which allow for growing good quality films with thickness down to the sub-100-nm range\cite{Wang2014,onbasli2014pulsed,sokolov2016thin,hauser2016yttrium,Cooper2017} and even below 10~nm\cite{sun2012growth,d2013inverse,krichevtsov2017magnetization}. Since the thickness as well as structural and compositional effects have a large influence on the magnetic behavior, these developments call for a detailed characterization of the magnetic properties of ultrathin YIG films. Several characteristic quantities, which are of high relevance for YIG-based spintronics, have been found to vary in films thinner than 100~nm. For instance, a reduction of the saturation magnetization\cite{popova2001perpendicular,onbasli2014pulsed,sokolov2016thin,hauser2016yttrium,Cooper2017,mitra2017interfacial,gomez2018synthetic} is typically observed in YIG films with thickness down to 10~nm, which has been ascribed to either thermally-induced stress\cite{popova2001perpendicular}, lack of exchange interaction partners at the interface\cite{sokolov2016thin}, or stoichiometric variations\cite{hauser2016yttrium,Cooper2017,mitra2017interfacial,gomez2018synthetic}. Furthermore, an increase of the damping\cite{sun2012growth,d2013inverse,onbasli2014pulsed,PhysRevB.91.134407} as well as a decreased spin mixing conductance\cite{PhysRevB.92.054437,PhysRevLett.115.096602} have been found in ultrathin YIG. Finally, the emergence of unexpected magnetocrystalline anisotropy was reported for films of different orientations grown on gadolinium gallium garnet (GGG) and yttrium aluminium garnet (YAG). The magnetic anisotropy was investigated by the magneto-optical Kerr effect in GGG/YIG(111)\cite{krichevtsov2017magnetization,sokolov2016thin} and by ferromagnetic resonance in GGG/YIG(111)\cite{manuilov2009pulsed}, GGG/YIG(001)\cite{manuilov2010pulsed}, and YAG/YIG(001)\cite{Wang2014}. Whereas all these studies address important magnetic characteristics in the sub-100 nm range, only few studies\cite{sun2012growth,d2013inverse} explore the ultrathin film regime below 10 nm. This thickness range is highly relevant for efficient magnetization manipulation using current-induced interfacial effects as well as for strain engineering, since strain and its gradients relax after 10 to 20 nm. Finally, a comprehensive knowledge of the domain and domain wall structure in the thin film regime is lacking. Recent studies on magnetic domains address only bulk\cite{Lisovskii2013}, several micrometers\cite{temiryazev2003surface} or hundreds of nanometers\cite{xia2010investigation} thick YIG films.

In this work, we present a systematic investigation of the structure, saturation magnetization, magnetic anisotropy, and magnetic domains of YIG/Pt films grown on GGG substrates by PLD as a function of YIG thickness from $t_{\text{YIG}}=3.4$ to $t_{\text{YIG}}=90$~nm. By combining x-ray diffraction (XRD), transmission electron microscopy (TEM), atomic force microscopy (AFM) and x-ray absorption spectroscopy, we show that our films possess high crystalline quality and smooth surfaces with no detectable interface mixing throughout the entire thickness range. The saturation magnetization, investigated using a superconducting quantum interference device (SQUID), shows values close to bulk for thick films and a gradual reduction towards lower thicknesses. We probe the magnetic anisotropy electrically by means of SMR and find an easy plane and uniaxial in-plane anisotropy with a non-monotonic variation of the in-plane orientation of the easy axis and the magnitude of the effective anisotropy field. Finally, we investigate the domain structure using x-ray photoelectron emission microscopy (XPEEM), evidencing significant changes in the domain structure above and below $t_{\text{YIG}}=10$~nm.
Our results provide a basis for understanding the behavior of spintronic devices based on YIG/Pt with different YIG thickness.

\begin{figure*}[hbtp!]
	\includegraphics[width=18cm]{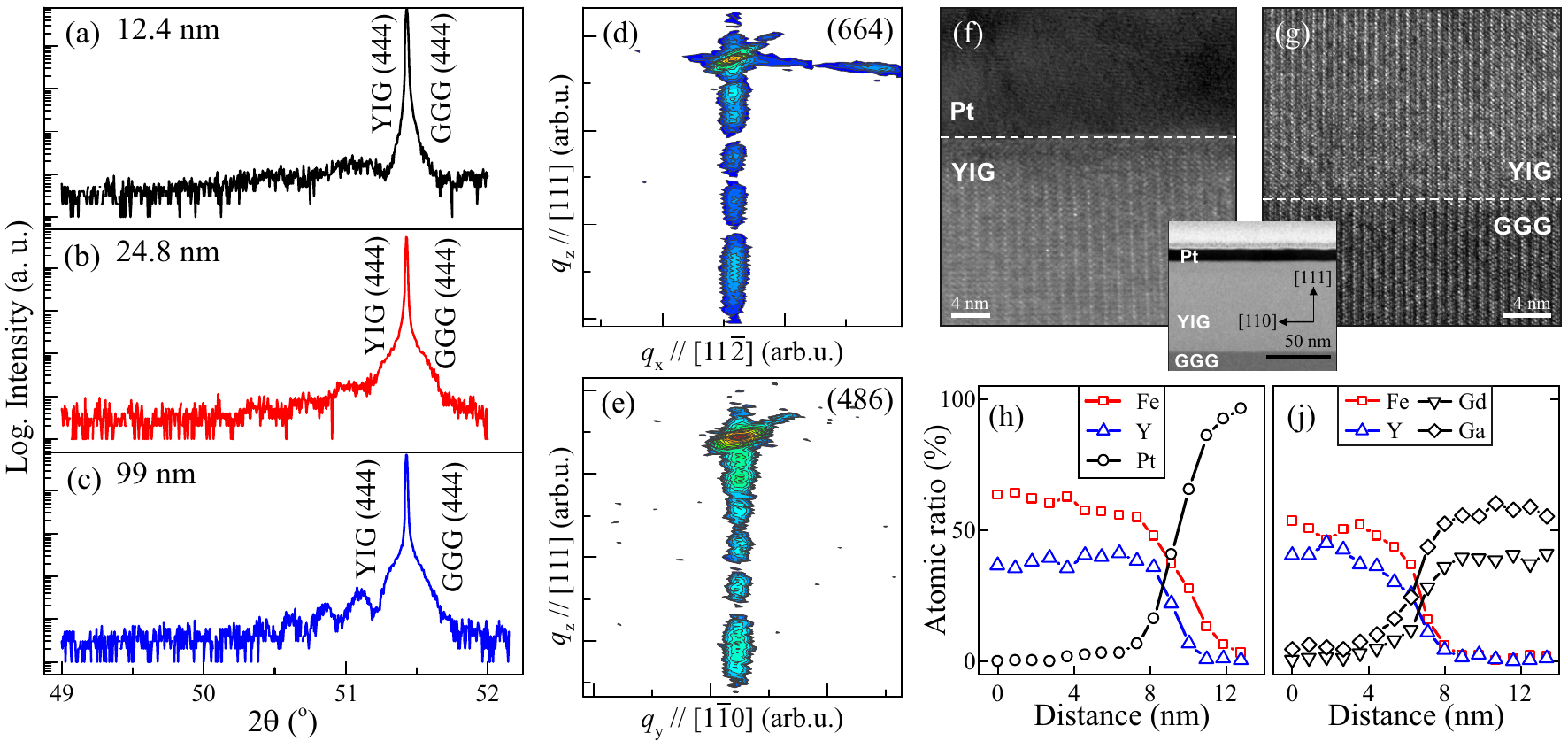}
	\caption{\label{fig:structure} (a-c) Structural characterization of GGG/YIG(111) films of different thickness by $\theta-2\theta$ scans using XRD. (d,e) Reciprocal space maps around the (664) and (486) diffraction peaks. The oscillations in the $\theta-2\theta$ and $q_{\rm z}$ scans are thickness oscillations. The thickness derived from these measurements agrees with that measured using XRD. (f,g) High resolution STEM image of the two interfaces in the stack GGG/YIG(72)/Pt(9). The inset shows a cross-section of the full stack. (h,j) Elemental profiles across the YIG/Pt and YIG/GGG interfaces measured by EDS.}
\end{figure*}

\section{Growth and Structure}
We prepared three sample series consisting of GGG/YIG($t_{\text{YIG}}$)/Pt bilayers with Pt thickness set to 3 nm in the first two series for optimal SMR measurements and 1.9 nm in the third series to allow for surface-sensitive XPEEM measurements while avoiding charging effects. The YIG thickness, $t_{\rm YIG}$, varies from 3.4 to 90~nm. An overview of the samples and their thickness in nm and YIG unit cells can be found in Table~\ref{tab:samples}. The samples have been grown in-situ on (111)-oriented GGG substrates using an ultra-high vacuum PLD system combined with dc magnetron sputtering (base pressure $10^{-8}$ mbar, $10^{-10}$ mbar, respectively). The YIG was grown by PLD at a growth pressure of $10^{-1}$ mbar and temperature of 720 $^\circ$C using an excimer KrF laser (wavelength 248 nm) at a repetition rate of 8 Hz and 1.45 J/cm$^2$ fluence. Reflection high energy electron diffraction (RHEED) was used to monitor the growth rate in-situ. After cooling down under a 200~mbar O$_2$ atmosphere, the GGG/YIG films were transferred to the sputtering chamber without breaking the ultra-high vacuum, where Pt was deposited at room temperature under $10^{-2}$ mbar Ar pressure. Finally, for electrical transport measurements, the YIG($t_{\text{YIG}}$)/Pt(3) series were patterned into Hall bar structures using optical lithography followed by Ar-ion milling. The Hall bars are oriented parallel to the $[1\bar{1}0]$ crystal direction and are 50 \textmugreek m wide, with a separation of the Hall arms of 500 \textmugreek m.
\begin{table}[hbtp!]
	\begin{ruledtabular}
		\begin{tabular}{lcccccccc}
			&$t_{\text{Pt}}$ [nm]&\multicolumn{7}{c}{$t_{\text{YIG}}$ [nm, (unit cells)]}\\
			\hline
		   SMR 1& 3 & 3.4,	& 4.6,	& 6.2,	& 7.3,	& 9,	& 29,	& 90\\
			    &   & (2.7) &(3.7)	& (5.0)	& (5.9)	& (7.3)	& (23.4)& (72.7)\\
			    \hline
		   SMR 2& 3 & 		&3.7, 	& 5.6,	& 6.2,	& 6.8,	& 7.4,	& 12.4\\
		   		&	&		&(3.0)	&(4.5)	&(5.0)	&(5.5)	&(6.0)	&(10.0)\\
			\hline
			XPEEM & 1.9& & &3.7,& 8.7,& 12.4,& 28.5,& 86.7\\
			& & & &(3.0)&(7.0)&(10.0)&(23.0)&(70.0)
		\end{tabular}
	\end{ruledtabular}
	\caption{\label{tab:samples}Overview of the three sample series used for anisotropy characterization using SMR and domain structure using XPEEM where $t_{\text{Pt}}$ and $t_{\text{YIG}}$ refer to the Pt and YIG thickness, respectively. The numbers between parentheses refer to the thickness in unit cells of YIG (1 unit cell=1.238~nm~[\onlinecite{GELLER195730}]).}
\end{table}

The crystalline quality and epitaxial strain of the GGG/YIG(111) films were investigated using XRD to obtain $\theta-2\theta$ diffraction scans and reciprocal space maps. The interface quality was probed using scanning TEM (STEM), energy dispersive x-ray spectroscopy (EDS), and electron energy loss spectroscopy (EELS) in order to resolve the elemental composition, whereas the thin film topography was investigated using AFM in both contact and tapping mode.

The YIG films are single phase epitaxial layers with (111) orientation, as indicated by the diffraction peaks corresponding to the (111) planes in the diffraction pattern. The x-ray diffractograms for three selected YIG thicknesses are shown in Fig. \ref{fig:structure} (a-c). Kiessig thickness fringes indicate the homogeneous growth and high interface quality. The observed periodicity matches the thicknesses determined by x-ray reflectivity. To further investigate the crystalline quality and the interfacial strain states of the YIG films, we recorded reciprocal space maps around the (664) and (486) diffractions for both YIG and the GGG substrate. These maps allow us to probe the strain state of the films along the two inequivalent [11$\bar{2}$] and [1$\bar{1}$0] in-plane directions. A coherent strain state of the films was evidenced for all thicknesses considered. The good lattice matching between YIG and GGG results in high quality epitaxial growth and no detectable strain relaxation, as shown by the alignment of the diffraction points along both $[11\bar{2}]$ and $[1\bar{1}0]$ orientations. The reciprocal space maps measured on a 99~nm-thick YIG film are shown in Fig. \ref{fig:structure} (d) and (e).

\begin{figure}[hb!]
	\includegraphics[width=0.8\columnwidth]{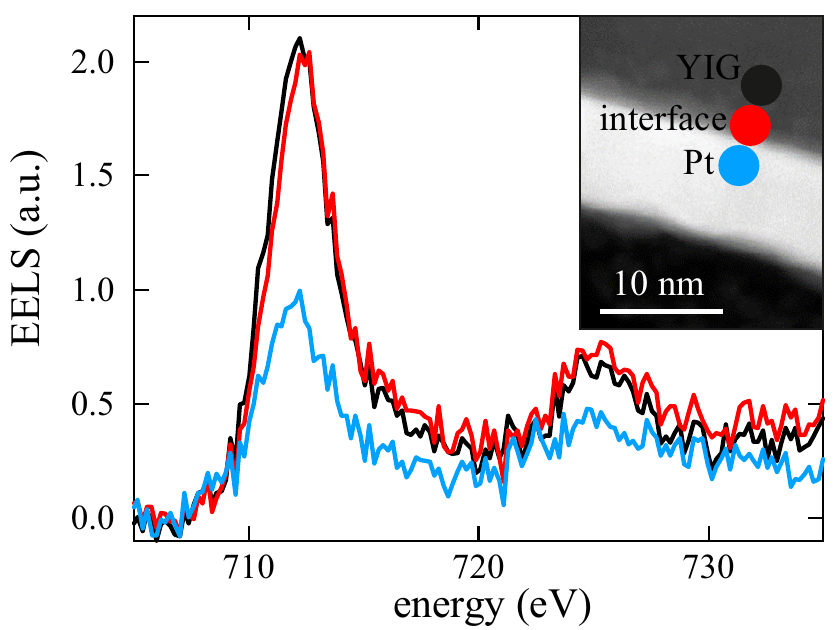}
	\caption{\label{fig:Fe_eels} Comparison of the Fe $L_{2,3}$ spectra measured by EELS across the YIG/Pt interface. The spectra, shown after background subtraction, have been averaged on 3~nm-long line scans for each region of interest, as labeled in the inset with colored dots. The probe beam diameter is 0.15~nm.}
\end{figure}

Figures~\ref{fig:structure} (f,g) show high resolution STEM images of the two interfaces and a TEM image of the full stack in the inset. The corresponding chemical profiles across the interfaces probed by EDS are shown in Figs.~\ref{fig:structure} (h,j). The EDS profiles indicate a moderate interdiffusion of Fe, Y, and Pt within a range of 2~nm at the YIG/Pt interface and of Gd and Ga into YIG within a range of about 4~nm at the GGG/YIG interface. Both values are in agreement with recent findings\cite{chang2017sputtering,mitra2017interfacial}. The smaller diffusion range at the YIG/Pt interface is consistent with the low power and lower deposition temperature (room temperature) of Pt compared to the deposition of YIG on GGG. The moderate diffusion of Ga and Gd into the YIG film is also in agreement with previous reports\cite{mitra2017interfacial,Cooper2017}.	
The EELS analysis, shown in Fig.~\ref{fig:Fe_eels}, confirms the moderate diffusion of Fe in the Pt.
%No Fe edge shift near the Pt interface could be observed. This can be due to our 1.4~eV energy resolution.
In order to gain information on the evolution of the local Fe environment across the YIG/Pt interface, we calculated the intensity ratio of the Fe $L_3$ and $L_2$ white lines from the Fe EELS spectra measured within the YIG film (black dot in Fig.~\ref{fig:Fe_eels}) and at two different positions near the YIG/Pt interface (red and blue dots). In YIG, the $L_3/L_2$ ratio is 5.64, which is consistent with an Fe$^{3+}$ oxidation state, as found, e.g., in Fe$_2$O$_3$ (Ref.~\onlinecite{Colliex1991EELS}). On the Pt side of the YIG/Pt interface, the $L_3/L_2$ ratio decreases to 4.24, which is in between the values found for Fe and FeO. This change of electronic valence confirms the presence of Fe in the interfacial Pt layer. We note that the absence of Fe$^{3+}$ in the Pt further excludes measurement artifacts due to sample preparation for TEM, such as Pt redeposition %(and not Pt redeposition on YIG).

\begin{figure}[hbtp!]
	\includegraphics{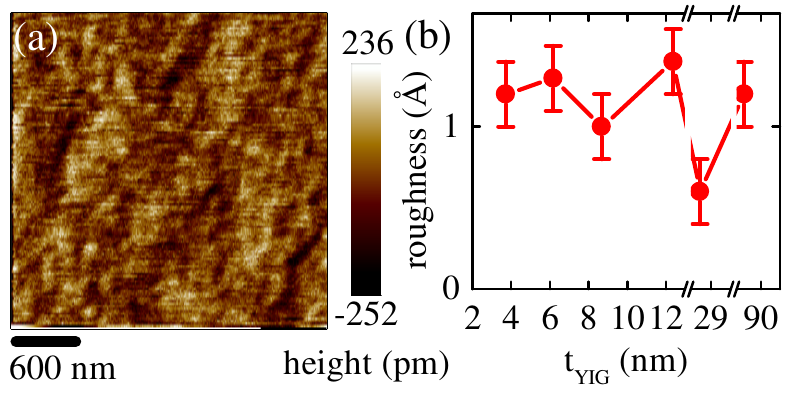}
	\caption{\label{fig:AFM} (a) AFM image of YIG(28.5)/Pt(3). (b) Root mean square roughness as a function of YIG thickness measured by AFM.}
\end{figure}

The AFM measurements substantiate the low roughness expected from the x-ray reflectivity and the observation of Kiessig fringes in the diffraction spectra. Figure \ref{fig:AFM} (a) shows an AFM image over an area of 3$\times$3 \textmugreek m$^2$ of a 28.5~nm thick GGG/YIG film. The measured roughness is in the range of 1 \AA\ and is independent of the YIG thickness, as shown in Fig. \ref{fig:AFM} (b).

\section{Saturation Magnetization}
\label{seq:mm}
\begin{figure}[hbtp!]
	\includegraphics{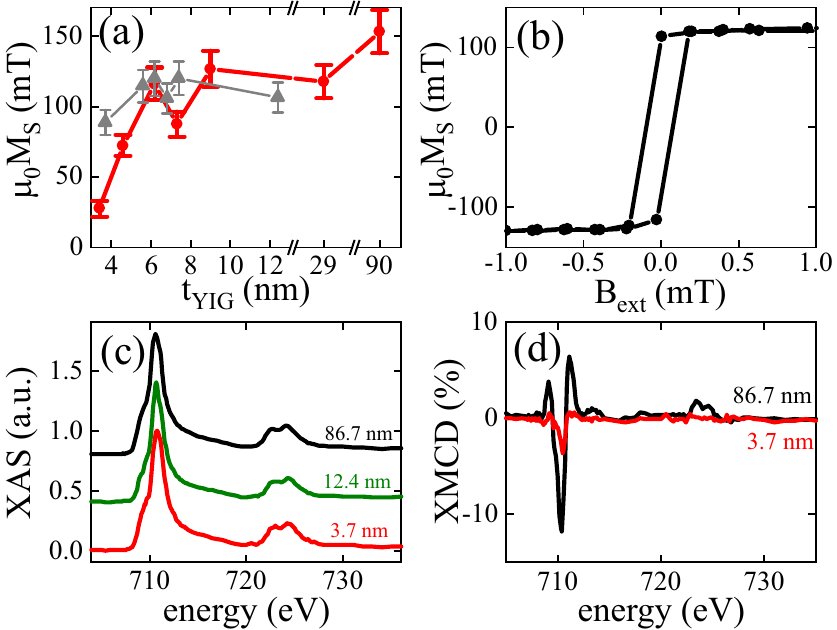}
	\caption{\label{fig:Ms} (a) $M_{\rm s}$ as a function of YIG thickness measured by SQUID for sample series SMR 1 (red) and SMR 2 (grey). (b) Magnetic hysteresis of YIG(9)/Pt(3) as a function of in-plane magnetic field. (c) X-ray absorption spectra of YIG(3.7,12.4,86.7)/Pt(1.9). Each line represents the sum of two spectra acquired with positive and negative circular x-ray polarization. The spectra are shifted by a constant offset for better visibility. (d) XMCD spectra for the thickest and thinnest samples obtained by taking the difference between two absorption spectra acquired with positive and negative circular x-ray polarization. The spectra were acquired on homogeneously magnetized domains in the XPEEM setup.}
\end{figure}

Figure \ref{fig:Ms} (a) shows the saturation magnetization ($M_{\rm s}$) of the unpatterned YIG($t_{\text{YIG}}$)/Pt(3) bilayers measured by SQUID (sample series SMR 1 and SMR 2 in red and grey, respectively). For each thickness, we observe a hysteretic in-plane magnetization with coercivity smaller than 0.5 mT, as illustrated for YIG(9)/Pt(3) in Fig. \ref{fig:Ms} (b). The thickest sample, YIG(90)/Pt(3), has a saturation magnetization of $\mu_0M_{\text{s}}=153$ mT, which is close to the bulk value of $\mu_0M_{\text{s,bulk}}=180$ mT at room temperature.\cite{hansen1974saturation} The reduction of $M_{\rm s}$ compared to the bulk was also observed in other studies of YIG films grown by PLD\cite{sokolov2016thin,popova2001structure,dumont2005superexchange,manuilov2009pulsed}. This reduction has been ascribed to the different percentage of Fe vacancies at the tetrahedral or octahedral sites\cite{manuilov2009pulsed}, the lack of exchange interaction partners for atoms at the interface\cite{sokolov2016thin}, strain relaxation due to a slight lattice mismatch of the substrate and YIG\cite{popova2001structure}, as well as to the presence of Fe and O vacancies\cite{dumont2005superexchange}. Furthermore, we observe a decreasing trend for $M_{\rm s}$ at lower YIG thicknesses, with a steeper reduction below 10 nm. Our thinnest sample ($t_{\text{YIG}}=3.4$ nm) has $\mu_0M_{\text{s}}=27$ mT, which is only 15\% of the bulk saturation magnetization. Previous studies reported a decreasing trend of $M_{\rm s}$ already for thicknesses larger than 10 nm (Refs.~\onlinecite{popova2001structure,mitra2017interfacial,Cooper2017,gomez2018synthetic,sokolov2016thin}).

The reduction of the YIG magnetization in thin films has been often modeled as a magnetically dead layer. For example, by extrapolating the areal magnetization as a function of thickness to the point where no surface magnetization would be present, Mitra et al.\cite{mitra2017interfacial} inferred a 6~nm thick dead layer for YIG films in the 10-50 nm thickness range. A similar extrapolation of our $M_{\rm s}$ data for samples with thickness of 9~nm and above would lead to a 4.3~nm thick dead layer. However, our data evidence a finite magnetization below 4.3~nm, contradicting the assumption of an abrupt magnetically dead layer at the GGG/YIG interface. Instead, we conclude that a gradual reduction of $M_{\rm s}$ occurs at thicknesses below 10~nm, which may be due to the diffusion of Gd from the GGG substrate into YIG observed by EDS [see Fig.~\ref{fig:structure}~(j)].

Further information on the magnetization of YIG($t_{\text{YIG}}$)/Pt(1.9) was obtained using x-ray absorption spectroscopy and x-ray magnetic circular dichroism (XMCD) at the $L_3$ and $L_2$ absorption edges of Fe. The x-ray absorption spectra of representative YIG(3.7,12.4,86.7)/Pt(1.9) samples present a very similar lineshape [Fig. \ref{fig:Ms} (c)], which implies that the chemical environment of the Fe atoms does not change substantially with thickness. The XMCD asymmetry, however, decreases in the thinner samples [Fig. \ref{fig:Ms} (d)], consistent with the behavior of $M_{\rm s}$ discussed above. These data confirm that our samples are magnetic throughout the entire thickness range and suggest that the reduction of $M_{\rm s}$ is not due to defective Fe sites.

\section{Magnetic Anisotropy}
\label{seq:ani}
To probe the magnetic anisotropy, we performed measurements of the transverse SMR\cite{nakayama2013spin,chen2013theory} as a function of magnitude and orientation of the external magnetic field $B_{\rm ext}$. Our data evidence the presence of an easy-plane anisotropy field, $B_{\text{K}_1}$, which adds to the demagnetization field to favor the in-plane magnetization, as well as of an easy axis field $B_{\text{K}_2}$, which favors a particular in-plane direction that varies from sample to sample.

\subsection{Transverse SMR measurements}
A sketch of the Hall bar structure employed for the SMR measurements is presented in Fig.~\ref{fig:OOP}~(a). We used an ac current $I=I_0\sin(\omega t)$, modulated at a frequency $\omega/2\pi=10$~Hz, and acquired the longitudinal and transverse (Hall) resistances\cite{garello2013symmetry,avci2015unidirectional}. To extract the magnetization orientation, we consider here only the transverse resistance, $R_{\text{xy}}$, which is sensitive to all three Cartesian components of the magnetization due to the SMR effect\cite{chen2013theory}. We performed two types of measurements. In the first type, which we call IP angle scan, we vary the in-plane angle of the applied magnetic field $B_{\rm ext}$; in the second type, which we call OOP field scan, we ramp $B_{\rm ext}$ applied out-of-plane. For both cases, it is convenient to use spherical coordinates: we define $\varphi_{\text{B}}$ as the azimuthal angle between $B_{\rm ext}$ and $I$ and $\varphi$ as the azimuthal angle between the magnetization and $I$, whereas the polar angles of $B_{\rm ext}$ and the magnetization with respect to the surface normal are $\theta_{\text{B}}$ and $\theta$, respectively [Fig.~\ref{fig:OOP}~(a)].

In the IP angle scan ($\theta=\pi/2$), $R_{\text{xy}}$ is determined solely by the planar Hall-like (PHE) contribution from the SMR, leading to
\begin{align}
	\label{eq:Rphe}
	R_{\text{xy}}=R_{\text{PHE}}\sin(2\varphi),
\end{align}
where $R_{\text{PHE}}$ is the planar Hall-like coefficient. In this type of experiment, we record $R_{\text{xy}}$ as a function of $\varphi_{\rm B}$ for different $B_{\text{ext}}$. For fields large enough to saturate the magnetization along the field direction, we can assume $\varphi=\varphi_{\text{B}}$ and hence $R_{\text{xy}}=R_{\text{PHE}}\sin(2\varphi_{\text{B}})$. Conversely, for external fields small enough such that $\varphi\neq\varphi_{\text{B}}$, $R_{\text{xy}}$ will deviate from the $\sin(2\varphi_{\text{B}})$ curve. By adopting a macrospin model assuming in-plane uniaxial magnetic anisotropy and comparing the resulting PHE from Eq.~(\ref{eq:Rphe}) with our data, we can determine quite accurately the easy axis direction as well as the magnitude of the in-plane magnetic anisotropy energy.

In the OOP field scans, $R_{\text{xy}}$ depends on the ordinary Hall effect (OHE) and anomalous Hall-like (AHE) contribution from the SMR, which is proportional to the out-of-plane component of magnetization\cite{chen2013theory}. We thus have
\begin{align}
	\label{eq:Rahe}
	R_{\text{xy}}=R_{\text{OHE}}B_{\text{ext}}\cos\theta_{\rm B} +R_{\text{AHE}}\cos\theta,
\end{align}
where $R_{\text{OHE}}$ and $R_{\text{AHE}}$ are the OHE and AHE coefficients, respectively. When ramping the OOP field, the contribution due to $R_{\text{OHE}}$ continuously increases, whereas the contribution due to $R_{\text{AHE}}$ saturates as the magnetization is fully aligned out-of-plane. The corresponding saturation field $B_{\text{s}}$ can be used to determine the out-of-plane magnetic anisotropy knowing the value of $M_{\rm s}$. All measurements were performed at room temperature using a current density in the low $10^5$ A/cm$^2$ range.

\subsection{Easy plane anisotropy}
\label{sec:aniOOP}
\begin{figure}[hbtp!]
	\includegraphics{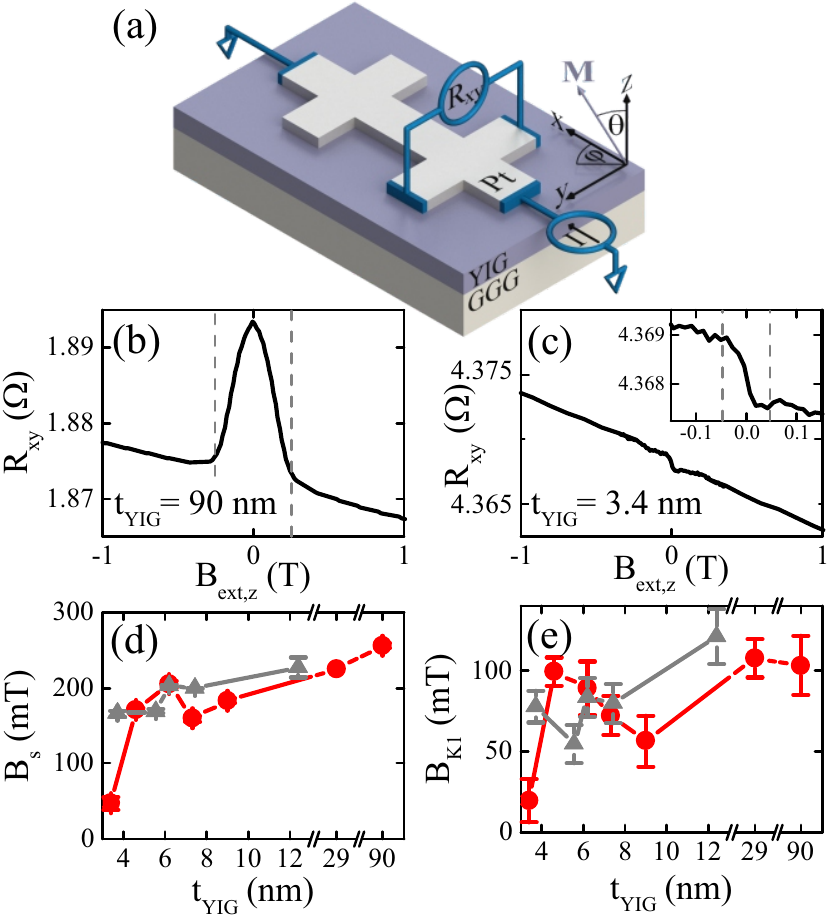}
	\caption{\label{fig:OOP} (a) Schematic of the Hall bar and coordinate system. OOP field scans for the thickest (90 nm) and thinnest (3.4 nm) YIG thickness from series SMR 1 are shown in (b) and (c), respectively. From these scans, the out-of-plane saturation field $B_{\text{s}}$ was determined as indicated by the dashed line and summarized in (d) as a function of YIG thickness. (e) Easy-plane anisotropy field $B_{\text{K}_1} = B_{\text{s}} - \mu_0M_{\rm s}$. Data shown for sample series SMR 1 (red) and SMR 2 (grey).}
\end{figure}

The easy-plane anisotropy field was determined by comparing the hard axis (out-of-plane) saturation field $B_{\text{s}}$ measured by the Hall resistance with the demagnetizing field $\mu_0M_{\rm s}$ estimated using SQUID. Figures \ref{fig:OOP} (b) and (c) show the results of OOP field scans for the thickest (90 nm) and thinnest (3.4 nm) YIG/Pt(3) samples. In both cases, we identify $B_{\text{s}}$ (dashed line) as the field above which only the ordinary Hall effect contributes to the (linear) increase of $R_{\text{xy}}$ with increasing field. Note that, for YIG(90)/Pt(3), we observe a bell-shaped curve that is due to the PHE during the re-orientation of magnetic domains from the initial in-plane to the final out-of-plane orientation at $B_{\text{s}}$. Figure \ref{fig:OOP} (d) reports the estimated values of $B_{\text{s}}$ for all thicknesses. In all cases, we find that $B_{\text{s}}$ is significantly larger than $\mu_0 M_{\rm s}$ reported in Fig.~\ref{fig:Ms} (a). We attribute this difference to an easy-plane anisotropy field, $B_{\text{K}_1} = B_{\text{s}} - \mu_0 M_{\rm s}$, which favors in-plane magnetization.
The magnitude of $B_{\text{K}_1}$ varies in the range of 50-100~mT (except for the thinnest sample). This additional easy plane anisotropy is of the same order of magnitude as in previous studies\cite{sokolov2016thin,krichevtsov2017magnetization}, where it was attributed to a rhombohedral unit cell distortion along the $[111]$ direction by $\approx$ 1 \%. Interestingly, the thinnest sample shows a drastically reduced $B_{\text{K}_1}$ compared to the thicker ones, suggesting that very thin YIG films with $t_{\text{YIG}}\leq 3.4$~nm tend to develop out-of-plane magnetic anisotropy\cite{popova2001perpendicular}. This tendency, however, is not sufficient to induce an out-of-plane easy axis, as verified by SQUID measurements as a function of out-of-plane field. The reduction of $B_{\text{K}_1}$ may also explain the absence of the bell-shaped PHE contribution to $R_{\text{xy}}$ observed in Fig. \ref{fig:OOP} (c).

   \begin{figure*}[hbtp!]
	\includegraphics{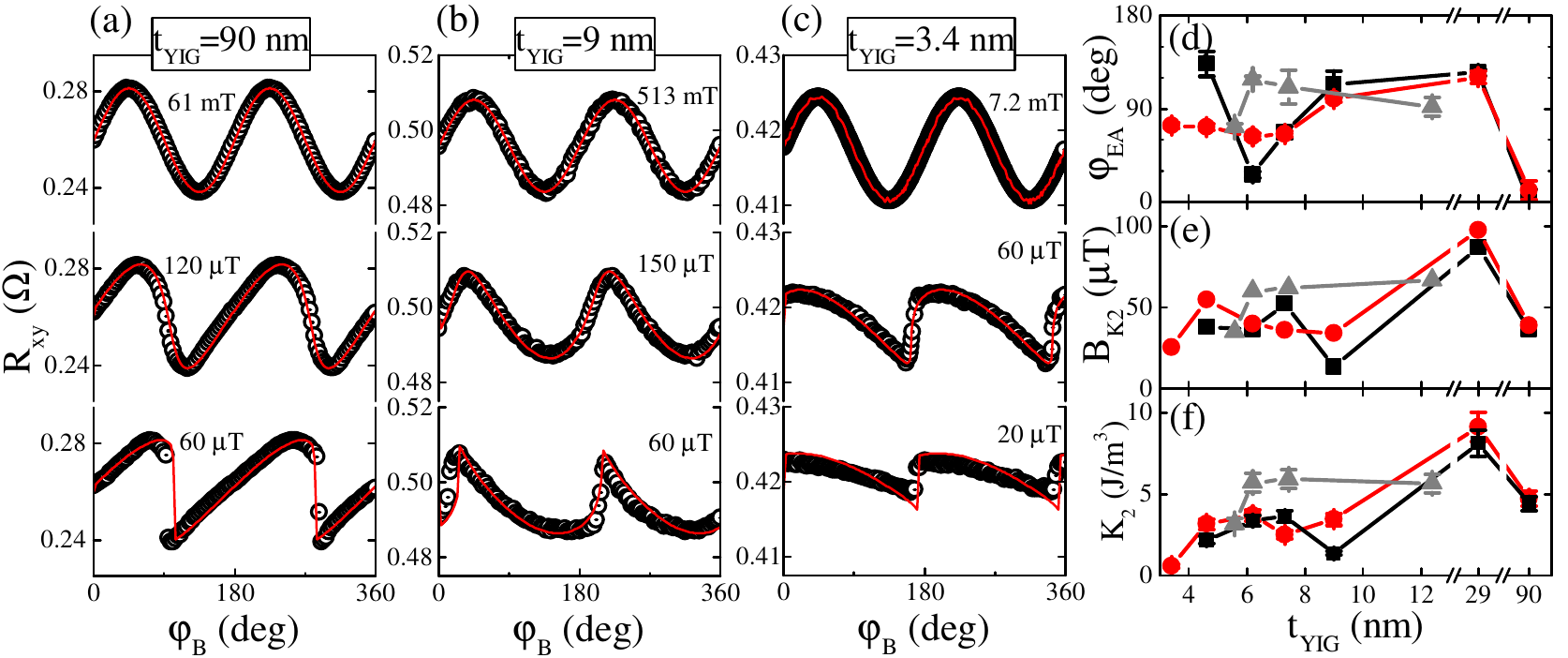}
	\caption{\label{fig:As_and_ani} Transverse resistance $R_{\text{xy}}$ as a function of azimuthal orientation of the external magnetic field, $\varphi_{\rm B}$, for different thicknesses in (a-c). The data are shown by black symbols, the macrospin simulations by red solid lines. (d) Easy axis orientation, $\varphi_{\rm EA}$, relative to the $[1\bar{1}0]$ crystal axis. (e) Effective uniaxial anisotropy field $B_{\text{K}_2}$ and (f) in-plane uniaxial energy $M_{\text{s}}B_{\text{K}_2}$ as a function of YIG thickness. Black and red points show the results for two different Hall bars patterned on the same chips of series SMR 1; grey triangles are results obtained on Hall bars of series SMR 2.}
\end{figure*}

\subsection{In-plane uniaxial anisotropy}	
The in-plane uniaxial anisotropy field $B_{\text{K}_2}$ is determined by measuring IP angle scans of $R_{\text{xy}}$ for different constant values of $B_{\text{ext}}$. By choosing $B_{\text{ext}}$ below the in-plane saturation field and fitting $R_{\text{xy}}$ as a function of $\varphi_{\rm B}$ using a macrospin model, we determine the orientation of the in-plane easy axis as well as the magnitude of $B_{\text{K}_2}$. The black circles in Fig. \ref{fig:As_and_ani} (a-c) show $R_{\text{xy}}$ as a function of $\varphi_{\rm B}$ for three representative YIG thicknesses (90, 9, 3.4 nm). For all three samples, we identify the $\sin(2\varphi)$ behavior expected from Eq.~(\ref{eq:Rphe}) for $B_{\text{ext}} > 7.2$~mT [upper panels in Fig. \ref{fig:As_and_ani} (a-c)]. When reducing $B_{\text{ext}}$, deviations from this lineshape occur, indicating that the magnetization no longer follows the external magnetic field. This behavior is most pronounced for fields of only tens of \textmugreek T [lower panels in Fig. \ref{fig:As_and_ani} (a-c)]. For such low fields, we observe two prominent jumps of $R_{\text{xy}}$ separated by 180$^{\circ}$, which we attribute to the magnetization switching abruptly from one quadrant to the opposite one during an IP angle scan in proximity to the hard axis. These jumps, which occur at different $\varphi_{\text{B}}$ for the three samples, indicate the presence of in-plane uniaxial anisotropy.

\begin{figure*}[hbtp!]
	\includegraphics{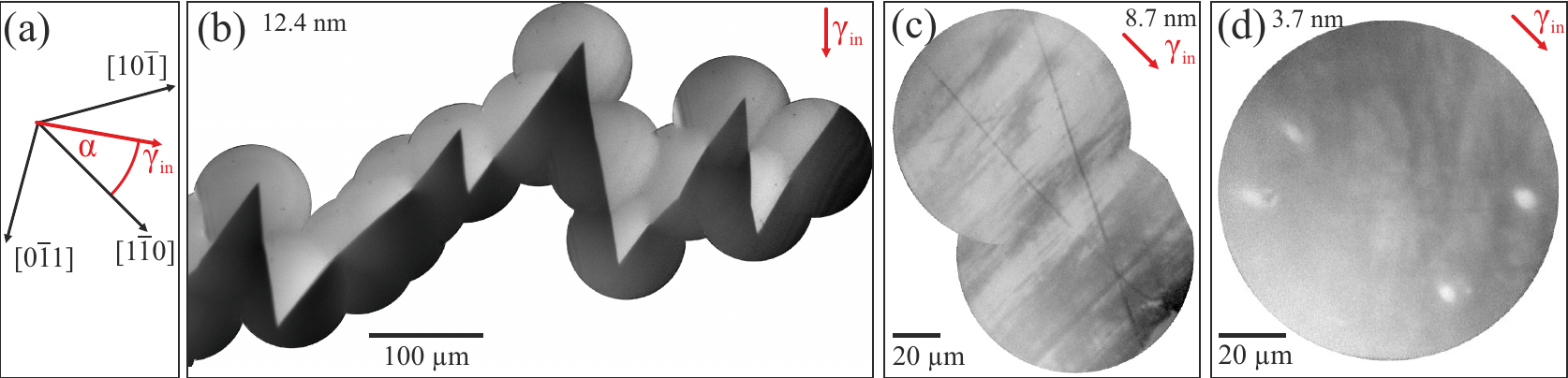}
	\caption{\label{fig:domain_structure} (a) Schematics of the direction of the incoming x-ray beam (red arrow) relative to the crystal axes in the XPEEM setup. (b-d) Domain structure of YIG/Pt(1.9) bilayers with $t_{\text{YIG}}=$ 12.4, 8.7, and 3.7~nm observed by XPEEM. The gray scale contrast corresponds to different in-plane orientations of the magnetization.}
\end{figure*}

In order to quantify the in-plane uniaxial anisotropy, we perform macrospin simulations of $R_{\text{xy}}$ assuming the following energy functional
\begin{align}
	E=-M_{\text{s}}\vec{m}\cdot\vec{B}_{\text{ext}}+M_{\text{s}}B_{\text{K}_2}\sin^2(\varphi-\varphi_{\rm EA}),
	\label{eq:Energy}
\end{align}
where $\vec{m}$ is the magnetization unit vector and $\varphi_{\rm EA}$ defines the angle of the easy axis with respect to the $[1\bar{1}0]$ crystal axis. Expressing $\vec{m}$ and $\vec{B}_{ext}$ in spherical coordinates, i.e., $\vec{m}=(\sin\theta\cos\varphi,\sin\theta\sin\varphi,\cos\theta)$ and $\vec{B}_{ext}=B_{ext}(\sin\theta_{\rm B}\cos\varphi_{\rm B},\sin\theta_{\rm B}\sin\varphi_{\rm B},\cos\theta_{\rm B})$, and taking $\theta = \theta_{\rm B}=\pi/2$, we can calculate the equilibrium position of the magnetization as a function of $B_{\rm ext}$, $\varphi_{\rm B}$, $B_{\text{K}_2}$, and $\varphi_{\rm EA}$. This calculation results in a set of values $\varphi(\varphi_{\rm B})$ that can be used to simulate $R_{\text{xy}}$ using Eq.~(\ref{eq:Rphe}). The planar Hall constants $R_{\rm PHE}$ required for the simulations are obtained by fitting Eq.~(\ref{eq:Rphe}) to the saturated data sets. Finally, we fit $R_{\text{xy}}(\varphi_{\text{B}})$ keeping $B_{\text{K}_2}$ and $\varphi_{\rm EA}$ as free parameters. The fits, shown as red lines in Fig. \ref{fig:As_and_ani} (a-c), reproduce the main features (lineshape and jumps) of $R_{\text{xy}}$ quite accurately, indicating that our method is appropriate to measure weak anisotropy fields in YIG. The resulting values of $\varphi_{\rm EA}$ and $B_{\text{K}_2}$ are shown by the symbols in Fig. \ref{fig:As_and_ani} (d) and (e), respectively.

The data for the series SMR 1 (red) in Fig. \ref{fig:As_and_ani} (d) appear to follow a pattern in the orientation of the easy axis as a function of thickness. Very thin samples with $t_{\text{YIG}}\leq 7.2$~nm have $\varphi_{\text{EA}}\approx60^\circ$, whereas intermediate thicknesses in the range $9\leq t_{\text{YIG}}\leq 29$~nm have $\varphi_{\text{EA}}\approx120^\circ$ and the thickest sample, $t_{\text{YIG}}=90$ nm, has $\varphi_{\text{EA}}\approx0^\circ$. While these orientations correspond to the three-fold symmetry of the (111) plane, measurements on different sets of samples show that this correspondence is likely coincidental. Measurements performed on devices patterned on the same chip (black squares) of series SMR 1 as well as on one device of the series SMR 2 (grey triangles) reveal uncorrelated variations of $\varphi_{\text{EA}}$ relative to the SMR 1 series, particularly for $t_{\text{YIG}} < 10$~nm. A variation of $\varphi_{\text{EA}}$ for samples grown on the same chip has also been observed in thicker YIG films\cite{krichevtsov2017magnetization} and is most likely due to inhomogeneities of process parameters such as, e.g., the temperature of the sample surface during deposition, which could lead to local strain differences.

Figure \ref{fig:As_and_ani} (e) shows $B_{\text{K}_2}$ as a function of $t_{\text{YIG}}$. Most values are close to 50~\textmugreek T, with minima and maxima of about 20 and 100 \textmugreek T, respectively. Combined with the thickness-dependent saturation magnetization from Fig. \ref{fig:Ms} (a), we obtain a magnetic anisotropy energy $K_2 = M_{\text{s}}B_{\text{K}_2}$ in the range of 0.1 to 10 J/m$^3$ [Fig. \ref{fig:As_and_ani} (f)]. $K_2$ is more than two orders of magnitude smaller than the first order cubic anisotropy constant of bulk YIG (-610 J/m$^3$) (Ref.~[\onlinecite{hansen1978rare}]), which reinforces the hypothesis of an extrinsic origin of the uniaxial anisotropy reported here.

Our observation of uniaxial anisotropy agrees with previous studies of YIG films grown by PLD on GGG\cite{PhysRevB.96.224403,manuilov2009pulsed,krichevtsov2017magnetization,sokolov2016thin}. From a crystallographic point of view, however, uniaxial anisotropy is not expected for the YIG(111) plane. Rather, for an ideal (111) crystal plane, one would expect a three-fold anisotropy due to the cubic magnetocrystalline anisotropy of bulk YIG\cite{CHUNG1975114}. This becomes obvious when translating the first order magnetocrystalline energy term of cubic crystals ($E_{\text{cubic}} \propto \alpha_1^2\alpha_2^2+\alpha_2^2\alpha_3^2+\alpha_3^2\alpha_1^2$, where $\alpha_{1,2,3}$ are the directional cosines with respect to the main crystallographic axes) into the coordinate system of the (111) plane ($\theta$ with respect to the [111] direction and, for simplicity, $\varphi$ with respect to the $[11\bar{2}]$ direction), giving
\begin{equation}
E_{\text{cubic}} \propto \frac{1}{4}\sin^4\theta+\frac{1}{3}\cos^4\theta +\frac{\sqrt{2}}{3}\sin^3\theta\cos\theta\cos 3\varphi . \label{eq:energy_3fold}
\end{equation}
In order for the threefold anisotropy to appear [last term in Eq.~(\ref{eq:energy_3fold})], the magnetization has to have a small out-of-plane component ($\theta\neq\pi/2$). Whereas this can be generally guaranteed in bulk crystals, in thin films the demagnetizing field and $B_{\text{K}_1}$ force $\theta = \pi/2$, resulting in the absence of threefold cubic anisotropy. The origin of the in-plane uniaxial anisotropy thus remains to be explained. Our XRD measurements reveal no significant in-plane strain anisotropy within the accuracy of the reciprocal space maps in Fig. \ref{fig:structure} (d) and (e). The small amplitude of $B_{\text{K}_2}$ and the intra- and inter-series variations of $\varphi_{\text{EA}}$ suggest that this anisotropy may originate from local inhomogeneities of the growth or patterning parameters, which should thus be taken into account for the fabrication of YIG-based devices.

\section{Magnetic domains}
\label{seq:PEEM}
The domain structure in bulk YIG is of flux-closure type with magnetic orientation dictated by the easy axes along the $[111]$-equivalent directions\cite{Basterfield1968,vlasko1976domain}. In the case of a defect-free YIG crystal, the domains can be as large as the sample itself, apart from the flux closure domains at the edges. Dislocations and strain favor the formation of smaller domains\cite{vlasko1976domain}. External strain and local stress due to dislocations can result in a variety of magnetic configurations such as cylindrical domains\cite{vlasko1976domain} and complex local patterns\cite{Basterfield1968,vlasko1974polarization}. Moreover, dislocations serve as pinning sites for domain walls\cite{vlasko1975features}. The domain walls observed at the surface of YIG single crystals are usually of the Bloch type and several $\mu$m wide \cite{Basterfield1968,Labrune1978}.

In the following, we characterize the domain structure of YIG thin films using XPEEM. The samples are YIG($t_{\text{YIG}}$)/Pt(1.9) bilayers with thickness down to $t_{\text{YIG}} = 3.7$~nm, as described for the XPEEM series in Table~\ref{tab:samples}. The measurements were performed at the SIM beamline of the Swiss Light Source\cite{le2012studying} by tuning the x-ray energy to obtain the maximum XMCD contrast at the Fe $L_3$ edge (710~eV) and using a circular field of view with a diameter of 100 \textmugreek m or 50 \textmugreek m. Magnetic contrast images were obtained by dividing pixel-wise consecutive images recorded with circular left and right polarized x-rays. Prior to imaging, the samples were demagnetized in an ac magnetic field applied along the surface normal.

\begin{figure*}[hbtp!]
	\includegraphics{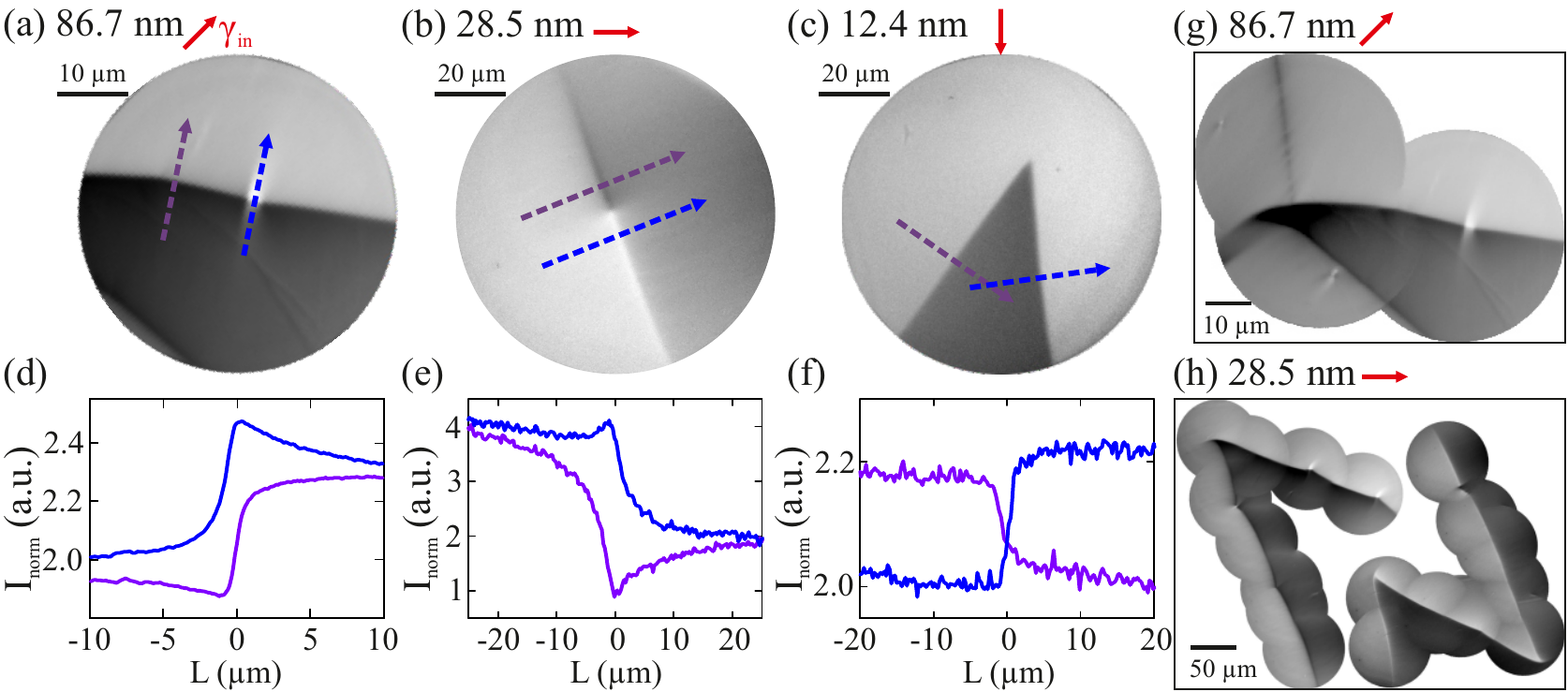}
	\caption{\label{fig:domain_details} XPEEM images of domain walls in (a) 86.7 nm, (b) 28.5 nm, and (b) 12.4 nm thick YIG. The arrows indicate the x-ray incidence direction. (d-f) Line profiles of the walls corresponding to the dashed lines in (a-c). (g,h) Details of the apices of the zigzag domains in 86.7 nm and 28.5 nm thick YIG. The samples are oriented with the crystal axes defined in Fig. \ref{fig:domain_structure} (a).}
\end{figure*}

\subsection{Domain structure}
Figure \ref{fig:domain_structure} shows representative XPEEM images of the magnetic domains of 12.4, 8.7, and 3.7 nm thick YIG. Thicker films ($t_{\text{YIG}}=$ 86.7, 28.5 and 12.4 nm) present qualitatively similar domain structures such as those shown in Fig.~\ref{fig:domain_structure} (b) for the 12.4 nm film. The domains in these thicker films extend over hundreds of \textmugreek m and meet head-on, separated by domain walls with characteristic zigzag boundaries. Such walls are typical of thin films with in-plane uniaxial anisotropy, where the zigzag amplitude and period depends on the balance between magnetostatic charge density and domain wall energy, as well as on the process of domain
formation\cite{Favieres_zigzagCo,hamzaoui1984static,hubert2008magnetic}. Note that the two sides of each zigzag are asymmetric, as found in ion-implanted garnets with uniaxial and trigonal in-plane anisotropy\cite{freiser1979zigzag}.
The length and opening angle of the zigzags increase and decrease, respectively, with increasing film thickness, similar to the trend reported for amorphous Ga-doped Co films with weak uniaxial anisotropy\cite{freiser1979zigzag}. In a quantitative model based on the mentioned energy balance\cite{labrune1982induced}, both quantities correlate with $M_{\text{s}}$ in a positive and negative manner, respectively. This trend is in agreement with the increase of $M_{\text{s}}$ reported in Fig. \ref{fig:Ms}~(a).

The domain morphology changes abruptly in films thinner than 12.4 nm. YIG films with  $t_{\text{YIG}}=8.7$ and 3.7 nm present much smaller domains, which extend only for tens of \textmugreek m and have a weaving pattern, as shown in Fig. \ref{fig:domain_structure} (c) and (d), respectively. In the 8.7 nm thick film, the domains elongate along the $[11\bar{2}]$ direction, likely due to the presence of in-plane uniaxial anisotropy. In the 3.7 nm thick film, the domains are more irregular and do not present such a strong preferred orientation, consistent with the reduced demagnetizing field and smaller in-plane anisotropy reported for this thickness [see Fig.~\ref{fig:OOP}~(e) and \ref{fig:As_and_ani} (f), respectively].

\subsection{Domain walls}
XPEEM images provide sufficient contrast to analyze the domain walls of films with $t_{\text{YIG}}\geq 12.4$ nm, which consist of straight segments along the zigzag boundary shown in Fig. \ref{fig:domain_structure} (b).
Figures~\ref{fig:domain_details} (a-c) show details of the $180^{\circ}$ domain walls that delimit the edges of the zigzag domains in 86.7, 28.5 and 12.4~nm thick YIG. Linecuts of the XMCD intensity across the walls are shown in  Figs.~\ref{fig:domain_details} (d-f). As the XMCD intensity scales with the projection of the magnetization onto the x-ray incidence direction (red arrows), the "overshoot" of the XMCD intensity in the central wall region compared to the side regions in Figs.~\ref{fig:domain_details}~(d,e) indicates that the magnetization rotates in the plane of the films. This observation is consistent with the N\'{e}el wall configuration expected for thin films with in-plane magnetic anisotropy\cite{hubert2008magnetic}, and contrasts with the Bloch walls reported at the surface of bulk YIG crystals\cite{Basterfield1968,Labrune1978}.
The walls appear to have a core region, corresponding to the length over which the magnetic contrast changes most abruptly, which is about 1-2~$\mu$m wide, and a tail that extends over several $\mu$m, which is a typical feature of N\'{e}el walls in soft films with in-plane magnetization\cite{hubert2008magnetic}.

The domain wall in Fig.~\ref{fig:domain_details} (b) shows an inversion of the magnetic contrast at a vortex-like singular point in the middle of the image. Such a contrast inversion reveals that the wall is composed by different segments
with opposite rotation mode of the magnetization, i.e., opposite chirality. The singular point that separates two consecutive segments is a so-called Bloch point, which can extend from the top to the bottom of the film, forming a Bloch line. Such features have smaller dimensions compared to the wall width and are therefore considered to favor the pinning of domain walls at defects, thus acting as a source of coercivity in soft magnetic materials\cite{hubert2008magnetic}.

Figures~\ref{fig:domain_details} (g) and (h) further show that the domain walls at the apices of the zigzag domains in the 86.7 and 28.5 nm thick films have a curved shape. This feature suggests that the domains are indeed pinned at defect sites, likely of structural origin \cite{vlasko1975features,vlasko1976domain}.
Finally, we note that the domain features reported here move under the influence of an external in-plane magnetic field of the order of few mT, as well as by ramping the out-of-plane field that compensates the magnetic field of the XPEEM lenses at the sample spot.

\section{Conclusions}
Our study shows that the magnetic properties and domain configuration of epitaxial YIG(111) films grown by PLD on GGG and capped by Pt depend strongly on the thickness of the YIG layer. Despite the high structural and interface quality indicated by XRD and TEM, the saturation magnetization decreases from $M_{\text{s}} = 122$~kA/m (15\% below bulk value) in 90~nm-thick YIG to 22 kA/m in 3.4~nm-thick YIG. The gradual decrease of $M_{\text{s}}$ suggests a continuous degradation of $M_{\text{s}}$ rather than the formation of a dead layer. All films except the thinner one ($t_{\text{YIG}}=3.4$~nm) present a rather strong easy-plane anisotropy in addition to the shape anisotropy, of the order of $10^3$ to $10^4$ J/m$^3$. Additionally, all films except the thinner one present weak in-plane uniaxial anisotropy, of the order of $3-10$~J/m$^3$. This anisotropy defines a preferential orientation of the magnetization in each sample, which, however, is found to vary not only as a function of thickness but also between samples with the same nominal thickness and even for samples patterned on the same substrate. The origin of this variation is tentatively attributed to local inhomogeneities of the growth or patterning process, which could lead to small strain differences not detectable by XRD. Besides these findings, we underline the fact that SMR measurements performed for external magnetic fields smaller or comparable to the effective anisotropy fields allow for the accurate characterization of both the magnitude and direction of the magnetic anisotropy of YIG/Pt bilayers, with an accuracy better than 10~\textmugreek T.

YIG films with $t_{\text{YIG}}=90-10$~nm present large, mm size, in-plane domains delimited by zigzag boundaries and N\'{e}el domain walls. The apices of the zigzags are pinned by defects, whereas the straight sections of the walls incorporate Bloch-like singularities, which separate regions of the walls with opposite magnetization chirality. The domain morphology changes abruptly in the thinner films. Whereas the 8.7 nm thick YIG presents elongated domains that are tens of \textmugreek m long, the domains in the 3.7 nm thick YIG are smaller and more irregular, consistent with the reduction of the easy-plane and uniaxial anisotropy reported for this sample. Our measurements indicate that the performance of YIG-based spintronic devices may be strongly influenced by the thickness as well as by local variations of the YIG magnetic properties.

\begin{acknowledgments}
We thank Rolf Allenspach for valuable discussions. Furthermore, we acknowledge funding by the Swiss National Science Foundation under Grant no. 200020-172775. Jaianth Vijayakumar is supported by the Swiss National Science Foundation through Grant no. 200021-153540. David Bracher is supported by the Swiss Nanoscience Institute (Grant no. P1502). Part of this work was performed at the SIM beamline of the Swiss Light Source, Paul Scherrer Institut, Villigen, Switzerland.
\end{acknowledgments}

\end{document}